\documentclass[aps,twocolumn,showpacs,preprintnumbers,amsmath,amssymb]{revtex4-1}
\usepackage{latexsym}
\usepackage{graphicx}
\usepackage{dcolumn}
\usepackage{bm}
\usepackage{xcolor}
\usepackage{empheq}
\usepackage{float}
\usepackage{enumitem}
\usepackage{amssymb,amsmath}
\usepackage{slashed}
\usepackage{hyperref}

\usepackage{CJK} 

\newcommand{\half}{{{\textstyle\frac{1}{2}}}}

\newcommand{\be}{\begin{equation}}
\newcommand{\ee}{\end{equation} }
\newcommand{\beqa}{\begin{eqnarray} }
\newcommand{\eeqa}{\end{eqnarray} }
\newcommand{\ba}{\begin{array}}
\newcommand{\ea}{\end{array}}

\newcommand\Tr{{\rm Tr}}

\newcommand\cC{{\cal C}}

\newcommand\cL{{\cal L}}

\def\i{\mathrm{i}}

\def\a{\alpha}

\def\g{\gamma}

\def\a{\alpha}
\def\b{\beta}

\def\l{{\lambda}}

\def\R{{L}}
\def\sc{{\phi}}
\def\l{\lambda}
\def\ve{\varepsilon}
\newcommand{\re}{{\rm e}}
\def\bar{\overline}

\newcommand{\nn}{\nonumber}
\renewcommand{\=}{\;  = \;}
\def\ads2{AdS$_2$}

\begin{document}
\begin{CJK}{UTF8}{mj}
\title{Supersymmetry and complexified spectrum on Euclidean AdS$_2$}
\author{Alfredo Gonz\' alez Lezcano} 
\email{alfredo.gonzalez@apctp.org}
\affiliation{
Asia Pacific Center for Theoretical Physics, Postech, Pohang 37673, Korea }

\author{Imtak Jeon} 
\email{imtakjeon@gmail.com}
\affiliation{
 Asia Pacific Center for Theoretical Physics, Postech, Pohang 37673, Korea }
\affiliation{
 Department of Physics, Postech, Pohang 37673, Korea 
}

\author{Augniva Ray}
\email{augniva.ray@apctp.org}
\affiliation{
Asia Pacific Center for Theoretical Physics, Postech, Pohang 37673, Korea }

\begin{abstract}
Quantum study of supersymmetric theories on Euclidean two dimensional anti-de Sitter space (AdS$_2$) is invalid if we use the standard normalizable functional basis due to its incompatibility with supersymmetry. We cure this problem by demonstrating that supersymmetry a requires complexified spectrum and constructing the  supersymmetric basis for scalar and spinor fields. Our new basis is free of fermionic zero modes, delta-function normalizable with respect to a newly defined inner product, and compatible with the  supersymmetric asymptotic boundary condition. We also explore the one-loop evaluation using this basis and show that it agrees with the standard nonsupersymmetric basis up to a global contribution arising from the fermion zero mode.

\end{abstract}
\maketitle
\end{CJK}

\section{Introduction}
Supersymmetric theories on Euclidean \ads2  have been an interesting subject of research in the last few decades. This is mainly because all extremal black holes  universally have an \ads2 factor in their near horizon geometry~\cite{Sen:2007qy}, where their thermodynamic properties  can be studied the via Euclidean path integral approach \cite{Gibbons:1976ue}. Supersymmetry  provides us with a powerful tool for studying quantum aspects of black hole entropy. 
For instance, the ``quantum entropy function" \cite{Sen:2008vm} defined as the Euclidean supergravity partition function  on their near horizon geometry in global coordinate for supersymmetric black holes has received a successful one-loop test by comparing it with corresponding microscopic results \cite{Sen:2011ba, Sen:2012kpz, Bhattacharyya:2012wz,Sen:2012cj, Sen:2012dw, Keeler:2014bra,Karan:2020njm, Banerjee:2021pdy}. 
Furthermore, computation using the supersymmetric localization method capturing all quantum corrections has achieved exact matching with microstate counting of black holes~\cite{Banerjee:2009af, Dabholkar:2010uh, Dabholkar:2011ec, Gupta:2012cy, Dabholkar:2014wpa, LopesCardoso:2022hvc, Hristov:2021zai, deWit:2018dix,Jeon:2018kec,Iliesiu:2022kny,Gupta:2021roy,Ciceri:2023mjl}.

However, despite all those extensive results,  quantum studies of supersymmetric theories on Euclidean \ads2 have suffered from a fundamental  underlying problem. (In the Lorentzian case~\cite{Sakai:1984vm}, no such problem occurs.) As noted in \cite{David:2018pex, David:2019ocd} and highlighted again in~\cite{Sen:2023dps}, quantum fluctuations of bosonic and fermionic fields are not mapped to each other by supersymmetry when expressed in terms of the standard delta-function normalizable eigenmodes of their respective kinetic operators \cite{Camporesi:1994ga, Camporesi:1992tm, Camporesi:1995fb
}, due to the asymptotic growth of Killing spinors. This would yield the conclusion  that supersymmetry demands non-normalizable modes over which the path integral would be ill defined, hence seemingly invalidating its evaluation exploiting supersymmetry.  
 This raises the following questions about the aforementioned quantum studies:  how is the localization method, which should rely on supersymmetry, valid and capable of giving the correct exact result, and how can the one-loop study using the heat kernel with standard nonsupersymmetric basis agree with supersymmetric results?
A recent thorough study on a simple theory on \ads2 has also supported  consistent results in both approaches~\cite{GonzalezLezcano:2023cuh}, yet an addressal of this question is awaited. 

We resolve this  problem by showing that supersymmetry on Euclidean global AdS$_2$ requires a complexified spectrum. We provide an explicit construction of the ``supersymmetric Hilbert space" for  scalar and spinor fields on that space by compexifying  the spectrum of the Dirac operator via the shift of the eigenvalue $\lambda$ by $\i/2$. With this construction, all the basis functions of the scalar are mapped by supersymmetry to the basis functions of spinor. These spectra form a Hilbert space, as they are still delta-function normalizable under an appropriate definition of the ``Euclidean inner product, " thereby  making the supersymmetric partition function well defined. Using an example of a supersymmetric theory, we demonstrate that this Hilbert space indeed provides a good supersymmetric basis to span all the fluctuations of fields demanded by the asymptotic boundary condition as dictated by the variational principle. 
The seemingly non-normalizable modes in the standard Hilbert space are now treated as normalizable modes in the supersymmetric Hilbert space. 
Additionally, we find that the new basis is free of fermion zero modes, making the path integral well defined. Using this basis, we explore the one-loop evaluation and show that the local part of the heat kernel computation is the same as the one obtained using the standard nonsupersymmetric basis. The difference in the global part arises from the fermion zero mode that can appear in the standard basis but not in the supersymmetric basis. Therefore, this construction provides a foundation for quantum study of supersymmetric theories on \ads2 \cite{David:2018pex, David:2019ocd,GonzalezLezcano:2023cuh,Cabo-Bizet:2017jsl}, thereby resolving the previously stated questions concerning supersymmetric black holes.  

\section{Delta-function normalizable modes}\label{deltaBasis}
We begin by revisiting the standard basis for scalar and spinor fields
given by delta-function normalizable eigenfunctions~\cite{Camporesi:1994ga, Camporesi:1992tm, Camporesi:1995fb
}  
of the Laplace and Dirac operators on the Euclidean global AdS$_2$ background whose metric is given by  
\begin{align*}
ds^2  =  \R^2(d\eta^2 + \sinh^2\eta\, d\theta^2)\,. 
\end{align*}
\paragraph{Scalar modes.$-$} 
For the Laplacian operator  $-\nabla^2$ on \ads2, the eigenfunctions are given by 
\be\ba{l}
\label{AdS2eigenfunction}
\phi_{\lambda, k}(\eta,\theta)=\frac{1}{\sqrt{2\pi }}\frac{1}{2^{| k|}| k|!} \!\left(\! \frac{\Gamma(\frac{1}{2}+| k| +\i\lambda)\Gamma(\frac{1}{2}+| k| -  \i \lambda )}{\Gamma(\i\lambda)\Gamma(-\i\lambda)} \right)^{\!\frac{1}{2}} \!\!{\rm e}^{\i k \theta}\\
\times {\sinh}^{\!| k|}\eta\,  F\!\left(\!\frac{1}{2}\!+\!| k|\!+\i\lambda, \!\frac{1}{2} \!+\!| k| \! - \!\i \lambda
 ; | k|\! +\!1; - \!\sinh^2\!\frac{\eta}{2} \right),
 \ea\ee
with $k \in \mathbb{Z}\,,~~\lambda \in \mathbb{R}_{>0}$,  where $F(\alpha, \beta; \gamma; z)$ is the hypergeometric function, which have eigenvalue, $L^{-2}(\lambda^2 +1/4)$. 
The eigenfunctions~\eqref{AdS2eigenfunction} satisfy delta-function orthonormality under the following definition of inner product (see Appendix \ref{Appendix:Innerproduct}):
\be\ba{l}\label{scalarinnerprod}
\big\langle \sc_{\lambda,k} | \sc_{\lambda',k'}\big\rangle  \!\equiv\! \int \! {\rm d}\eta {\rm d}\theta \sqrt{g}\,\sc_{\lambda,-k}
\sc_{\lambda',k'}
\!=\! L^2\delta(\lambda \!-\!\lambda')\delta_{k,k'}\,.
\ea\ee
Note that since we will complexify the parameter $\l$, we have defined the dual of a basis element $\sc_{\l, k}$ for the inner product as $\sc_{\l, -k}$, without using complex conjugation. This definition stems from the property that 
$\sc_{\lambda , - k}  = (\sc_{\lambda ,  k})^\ast \,
$
for real $\lambda$. 

Homogeneity of \ads2 implies that the spectral density can be obtained using the eigenfunctions evaluated at $\eta=0$, where only the $k=0$ mode survives. Hence, 
\be\ba{l}
\mu_{\sc}(\lambda)\equiv \sum_k \bigl( \sc_{\lambda,-k}\,\sc_{\lambda,k}\bigr)\,\bigr|_{\eta=0}\= \frac{\lambda}{2\pi}\tanh \pi \lambda \,.
\ea\ee
We note that $\l =0$ is not a part of the scalar spectrum since $\mu_\phi (0) =0$.

Using the inversion formula of hypergeometric function~(\ref{Inverse2F1}), one can show that  
the eigenfunctions have the following asymptotic behavior as $\eta \rightarrow \infty$
\be \label{scalarasymp}\ba{l}
\sc_{\lambda ,  k}(\eta, \theta)
\;\sim\;  {\rm e}^{-\frac{1}{2}\eta + \i k\theta}  \left( \a_{\lambda,k} {\rm e}^{\i \lambda\eta }+ \a_{-\lambda,k} {\rm e}^{-\i \lambda \eta } \right)\,,
\\
~~~~~~~~\a_{\lambda,k}\equiv \frac{1}{\pi \sqrt{2}}\left( \frac{\Gamma(\i\lambda)\Gamma(\half +|k| -\i\lambda )}{\Gamma(-\i\lambda)\Gamma(\half +|k| +\i\lambda )} \right)^{\!\frac{1}{2}}\,.
\ea\ee
\paragraph{Spinor modes.$-$} 
For the Dirac operator $\i \slashed{D}
$ on \ads2,  
with the gamma matrices being Pauli matrices, the  eigenfunctions are  given by \footnote{To map our notation to that used in \cite{GonzalezLezcano:2023cuh}, where the fermionic basis is organized using spinors $\chi^{\pm}_{\l. k}$ and $\eta^{\pm}_{\l,k}$, we have 
$\psi^+_{\pm|\l| ,k}  = \eta^\mp_{|\l|,k} $ and $
\psi^-_{\pm|\l| ,k} = \pm\chi^{\mp}_{|\l|,k} \,$ 
}
{
\begin{widetext}
\be \ba{l}\label{EigenAdS2a} 
\psi^+_{\l, k}=\! \frac{1}{\sqrt{4\pi }}  \frac{1}{k !} \! \left( \frac{\Gamma(1+k+\i\lambda) \Gamma(1+k -\i \lambda)}{\Gamma(\frac{1}{2} +\i\lambda)\Gamma(\frac{1}{2}-\i\lambda)}\right)^{\!\frac{1}{2}} \re^{\i(k+\frac{1}{2})\theta} \!
\begin{pmatrix}   \cosh^{k+1}\frac{\eta}{2}\sinh^{k}\frac{\eta}{2}\, F(k\!+\!1\!+\!\i\lambda, k\!+\!1\!-\!\i\lambda; k\!+\!1;-\sinh^2\!\frac{\eta}{2})\\ 
 -\i\frac{\lambda}{k+1} \cosh^{k}\!\frac{\eta}{2}\sinh^{k+1}\!\frac{\eta}{2}\,F(k\!+\!1\!+\!\i\lambda, k\!+\!1\!-\!\i\lambda; k\!+\!2; -\sinh^2\!\frac{\eta}{2})\end{pmatrix}, 
\\ 
\psi^-_{\lambda,k}= \! \frac{1}{\sqrt{4\pi }} \frac{1}{k !}\!\left( \frac{\Gamma(1+k+\i\lambda) \Gamma(1+k -\i \lambda)}{\Gamma\left(\frac{1}{2} +\i\lambda\right)\Gamma\left(\frac{1}{2}-\i\lambda\right)}\right)^{\!\frac{1}{2}}\! {\rm e}^{-\i(k+\frac{1}{2})\theta} 
\!\begin{pmatrix}\i \frac{\lambda}{k+1}  \cosh^k\frac{\eta}{2}\sinh^{k+1}\!\frac{\eta}{2}\,F(k\!+\!1\!+\!\i\lambda, k\!+\!1\!-\!\i\lambda; k\!+\!2;-\sinh^2\frac{\eta}{2})\\  -\cosh^{k+1}\frac{\eta}{2}\sinh^{k}\!\frac{\eta}{2}\,F(k\!+\!1\!+\!\i\lambda, k\!+\!1\!-\!\i\lambda; k\!+\!1; -\sinh^2\frac{\eta}{2})\end{pmatrix},
\ea\ee
\end{widetext}}
with $ \lambda \in \mathbb{R}\,,~~ k\in \mathbb{Z}_{\geq 0}\,,$ which have the eigenvalue, $L^{-1}\lambda \,$. 
The eigenfunctions~\eqref{EigenAdS2a}  satisfy delta-function orthonormality under the following definition of inner product:
\be\ba{l}\label{innerprodfermion}
\bigl\langle \psi^\pm_{\lambda,k} | \psi^\pm_{\lambda',k'}\big\rangle \equiv \pm \i \int {\rm d}\eta {\rm d}\theta \sqrt{g} \,  \psi^\mp_{\lambda,k} \,\psi^\pm_{\lambda',k^\prime} \\
~~~~~~~~~~~~~~~~=\!L^2\delta(\lambda\! -\!\lambda')\delta_{k,k'}\,,
\ea\ee
where our convention for the spinorial multiplication  is $\psi \chi \equiv \psi^T C \chi $ with $C =\g_2\,$. 
Note that we have defined the dual of a basis element $\psi^{\pm}_{\l, k}$  for the inner product through the symplectic Majorana conjugate as $\pm \i (\psi^{\mp}_{\l,k})^T C$, not using Hermitian conjugate. This definition stems from the property that 
$
(\psi^\pm_{\lambda,k})^\dagger 
= \pm \i (\psi^\mp_{\lambda,k})^T C\,
$
for real $\lambda$. 
In fact, this is the natural definition in Euclidean space  because Euclidean space treats the conjugate of a spinor as an independent spinor, formally doubling  fermionic degrees of freedom~\cite{Osterwalder:1972vwp,Osterwalder:1973zr}. Therefore, we would call \eqref{innerprodfermion}  ``Euclidean inner product."

Homogeneity of \ads2 implies that the spectral density can be obtained using the eigenfunctions evaluated at $\eta=0$, where only the $k=0$ mode survives. Hence, 
\be\ba{l}\label{spectralF}
\mu_{\psi^\pm}(\lambda) \equiv   \pm \i \sum_k  {\psi_{\lambda, k}^\mp} \,\psi_{\lambda,k}^\pm \,\Bigr|_{ \eta=0} =\frac{1}{4\pi} \lambda \coth\pi\lambda \,.
\ea\ee
We note that, unlike the case for scalar modes, the spectrum exists at $\lambda=0$ since $\mu_{\psi^{\pm}}(0) \neq 0 $. 

Finally, we note that the asymptotic behavior for large $\eta$ as follows:  
\be\ba{l} 
\psi^{\pm}_{\l,k} \!\sim  {\rm e}^{- \frac{\eta}{2}\pm \i(k+\frac{1}{2})\theta}   \bigl( {\rm e}^{\i \lambda \eta} \b_{\lambda,k} \upsilon_{(\!-\!)}\!  \pm {\rm e}^{-\i \lambda \eta} \b_{-\lambda,k} \upsilon_{(\!+\!)}\bigr) ,
\\
\label{alphaupsilon}
\beta_{\lambda,k}\equiv  \frac{1}{2\pi} 
\biggl(\frac{\Gamma\left(\frac{1}{2}+\i \l\right)\Gamma\left(1+k-\i \l\right)}{\Gamma\left(\frac{1}{2}-\i \l\right)\Gamma\left(1+k+\i \l\right)}\biggr)^{\!\frac{1}{2}}  , ~\upsilon_{(\pm)} \equiv \biggl( \ba{c} 1 \\  \pm1  \ea \biggr)\,,
\ea\ee
where we note the projection property, $P_\pm \upsilon_{(\pm)}= \upsilon_{(\pm)}$, with the projector, $P_\pm \equiv \half (1 \pm \gamma_1)$.

\section{Problem  with supersymmetry} \label{sec:asymp}
Let us elaborate on the problem concerning the supersymmetry of the standard delta-function normalizable modes. 
The supersymmetry relation between boson $\Phi$ and fermion $\Psi$ is generically given by local transformation using Killing spinors, $\varepsilon$,  as 
\be\ba{l}\label{genericsusy}
Q\Phi = {{\ve}} \Psi\,.
\ea\ee
On \ads2, 
 the Killing spinors satisfy the following  conformal Killing spinor equations, 
\be\ba{l}\label{KSEAdS2}
D_\mu \ve^s =   s\frac{1}{2L} \gamma_\mu  \ve^s\,,\qquad s=\pm 1\,,
\ea\ee
depending on the sign factor $s$ associated with the background value in gravity multiplet fields. 
With our choice of  the gamma matrices,  the Killing spinor solutions are 
\be\ba{l} \label{eq:KillingSpinors}
 {\ve}^s_+ \!= \!{\sqrt{L}}{\rm e}^{ \tfrac{\i\theta}{2}}\Biggl( \!\ba{c} \cosh \tfrac{\eta}{2} \\ \!s \sinh \tfrac{\eta}{2} \ea\!\!\Biggr), \, \, 
 {\ve}^s_- \!= \!{\sqrt{L}}{\rm e}^{- \tfrac{\i\theta}{2}}\Biggl(\!\ba{c} \!\!s \sinh \tfrac{\eta}{2}\\\cosh \tfrac{\eta}{2} \ea  \!\!\Biggr).
\ea\ee
We note that these spinors exponentially grow as $\exp({\eta /2})$  for large $\eta$, which  we call degree of growth $1/2$.

Let us look at the supersymmetry relation between scalar and Dirac spinors, for the left-hand and right-hand side of \eqref{genericsusy} respectively, in terms of the basis functions presented in previous section\ref{deltaBasis}. 
On one hand,  we recall from~\eqref{scalarasymp} that the basis functions for scalar have 
degree of growth $-1/2$.
On the other hand,  since the Killing spinors on the AdS$_2$ presented in \eqref{eq:KillingSpinors} have degree of growth $1/2$,  one can show using~\eqref{alphaupsilon} that the basis functions for spinor fields combined with the Killing spinors have degree of growth $0$ having the following asymptotic behavior, up to proportionality factors,  
\be\ba{ll}\label{asympbifer}
{{\ve}^s_\pm} \psi^\pm_{\lambda,k} \sim {\rm e}^{ s \i \lambda \eta \pm \i (k+1)\theta}\,,\qquad&
{{\ve}^s_\mp}\psi^\pm_{\lambda,k} \sim {\rm e}^{s \i \lambda \eta \pm \i k\theta}\,.
\ea\ee
 Therefore, the degree of growth of left- and right-hand sides of the supersymmetry relation~\eqref{genericsusy} do not match 
 when expressed in terms of the delta-function normalizable basis functions given in~\eqref{AdS2eigenfunction} and \eqref{EigenAdS2a}.

The mismatch at the level of asymptotic growth of basis elements is an indication that there is no mapping between the boson and fermion. To elaborate the argument, we try to find the supersymmetry transformation in terms of the mode expansion coefficient. 
If we schematically expand the boson and fermion in terms of complete basis $\sc_m$ and $\psi_n$, respectively, as $\Phi = \sum a_m \sc_m \,,   \Psi =\sum b_n \psi_n \,,$
then supersymmetry relates the bosonic coefficient  $a_m$ and the fermionic coefficient $b_m$  as
\be\ba{l}\label{susyofCoeff}
Q a_m = \langle \sc_m | {\ve}\Psi \rangle = \sum_n  b_n \langle \sc_m |  {{\ve}}\psi_n \rangle\,.
\ea\ee
Here, there is an issue: according to the asymptotic behavior of scalar in \eqref{scalarasymp} and bifermion in \eqref{asympbifer} having degree of growth $-1/2$ and $0$ respectively, the inner product $\langle \;\cdot\; | \;\cdot\;\rangle$ in \eqref{susyofCoeff} is ill defined as the integration diverges 
\be\ba{l}
\bigl\langle \sc_{\lambda',k'} \,|  {{\ve}}\psi_{\lambda, k} \bigr\rangle 
\sim \int_0^\infty {\rm d}\eta \sqrt{g}\,{\rm e }^{-\frac{1}{2}\eta} \rightarrow \infty \,.
\ea\ee
Nevertheless, we can  extract finite information by analytically continuing the integration. We  introduce the parameter $\epsilon >1/2$ such that  the integrand, including the measure $\sqrt{g},$  has degree of growth less than zero behaving as  $\exp[({1}/{2}-\epsilon)\eta] $ for large~$\eta$. Then we can perform the well defined integration,  
from which we  take the  $\epsilon$ to be zero.  
In this way one can show that the result is 
\be\ba{l}\label{ResultoScalarBifer}
\!\!\!\bigl\langle \sc_{\lambda^\prime,k} | {\ve^s} \psi_{\lambda,k} \bigr\rangle \propto  \delta\! \left( \lambda^{\prime} \!+\! \bigl( \lambda\! -\! s \frac{\i}{2}\bigr)  \!\right)\! +\!\delta\! \left( \lambda^{\prime} \!-\! \bigl( \lambda \!-\!  s\frac{\i}{2}\bigr)  \!\right).
\ea\ee
We shall omit the detail on how to introduce the parameter $\epsilon$ and perform the integration as this result is clear once we note the property of the bispinor that will be given in~\eqref{eq:bispinor}.
 
Since the spectral parameter 
for scalar and fermion basis is real, the inner product \eqref{ResultoScalarBifer} between the scalar basis and the bifermion  vanishes. This  result concludes that the superpartner of the scalar mode~\eqref{AdS2eigenfunction} is not expressible in terms of standard delta-function normalizable basis for spinor field~\eqref{EigenAdS2a}.
This seems to suggest that we have to give up the delta-function normalizable basis and find a way to introduce non-normalizable modes as argued in \cite{David:2018pex, David:2019ocd, Sen:2023dps}. However, in the following, we will show that we can still have the supersymmetric delta-function normalizable basis furnishing a supersymmetric Hilbert space.


\section{Supersymmetric Hilbert space with complexified spectrum}
In order to have  supersymmetry,  the inner product in~\eqref{ResultoScalarBifer} should  be nonzero.  From the  expression, we notice that it is natural to consider complexifying the parameter $\lambda$: if we consider fermionic modes with complex eigenvalue $\lambda$ by shifting with imaginary value $\i/2$, i.e. $\lambda \rightarrow \lambda + s \,\i/ 2$, then we obtain   $\delta(\lambda' -\lambda)$ in \eqref{ResultoScalarBifer}.
This idea of  shifting $\lambda$ by ${\i}/{2}$ is also supported by the following observation.
Using the Killing spinor equation~\eqref{KSEAdS2}, eigenvalue equation for $\psi^{\pm}_{\l , k}$, 
 and the fact that the scalar curvature of \ads2 is $R=-2 L^{-2}$, we can easily find the eigenvalue of bispinor with respect to the Laplace operator as
 \be\ba{l}
-\nabla^2 ({\ve^s} \psi^\pm_{\lambda,k})= \frac{1}{L^2}\left( \bigl(\l - s\frac{\i}{2}\bigr)^2 +\frac{1}{4} \right)({\ve^s} \psi^\pm_{\l,k}) \,.
\ea\ee
Therefore, if we shift the $\lambda \rightarrow \lambda +s  \, {\i}/{2}$, then the bispinor has the same eigenvalue of the scalar modes. This means that the bispinor is expanded in terms of the scalar eigenfunctions and the map between the scalar and fermion is  straightforward. In fact, one can show using the recursion relations \eqref{eq:gaussrecursion} that the bispinor with shifted $\lambda $ by $\i/2$ is exactly proportional to the eigenfunctions for scalar as \footnote{
If we use the Killing spinors that satisfy $D_\mu {\ve}^s =   s \i / 2\, \gamma_\mu \gamma_3 {\ve}^s$, then the following bispinors ${\varepsilon^{s}}^T C {U_s}\psi^\pm_{\lambda+  \i /2,k} $ are proportional to the scalar basis functions, where $U_s={\rm diag}(1, -s\i)$. 
}
\be\ba{ll} \label{eq:bispinor}
{\ve}^s_\pm \psi^\pm_{\lambda + s\frac{\i}{2},k} \propto 
\sc_{\lambda,\pm(|k|+ 1)}   \,, \, &
{\ve}^s_\mp \psi^\pm_{\lambda + s\frac{\i}{2},k} \propto 
\sc_{\lambda, \pm |k|}   \,.
\ea\ee

With the above idea,  we propose  a supersymmetric Hilbert space, which is composed of  the bosonic modes given in~\eqref{AdS2eigenfunction} together with the fermionic modes given in~\eqref{EigenAdS2a} 
with shifted $\lambda$ as 

\begin{empheq} [box=\fbox]{equation}
\ba{lll} \label{shiftedeigenfunc}
\mbox{ Scalar:  }&\Bigl\{ \sc_{\lambda, k}(\eta, \theta)  \, \Big|\, \lambda \in \mathbb{R}_{>0}\,, ~~k\in \mathbb{Z} \Bigr\}\,
\\
\mbox{ Spinor:  }&\Bigl\{\,\psi^\pm_{\lambda + s\frac{\i}{2},k}(\eta, \theta)\, \Big| \, \lambda \in \mathbb{R} \,, ~~k \in \mathbb{Z}_{\geq 0}\,\Bigr\}\,.
\ea\end{empheq}
 As a result,  all the modes of the scalar fields have their superpartners through the supersymmetry transformation~\eqref{genericsusy}, which is clear  from the relation~\eqref{eq:bispinor}.

The immediate consequence of the shifted $\lambda$ is that the fermionic modes have complex eigenvalue for the Dirac operator as $L^{-1} \left( \l + s \, \i / 2 \right)$.
We note that this complex eigenvalue is not unnatural because \ads2 is a noncompact space with boundary and the Dirac operator $\i\slashed{D}$ is no longer  Hermitian on that space. This non-Hermiticity of the Dirac operator was reported also in flat space with boundary~\cite{Bonneau:1999zq}.
By this shift, the spectral density of the spinor basis in supersymmetric Hilbert space now becomes,
\be\ba{l} \label{shifteddensity}
\mu_{\psi^{\pm}} (\lambda + s\frac{\i}{2}) \=  \frac{1}{4\pi}\Bigl(\lambda+ s\frac{\i}{2}\Bigr)\tanh\pi \lambda 
\,.
\ea\ee
We note that the  measure vanishes at $\lambda=0$ unlike the original measure in \eqref{spectralF}. 
 This implies that, while the mode with $\lambda=0$ in the basis given in \eqref{EigenAdS2a} is zero mode for massless case, the supersymmetric basis defined in~\eqref{shiftedeigenfunc} does not have such zero modes irrespective of the value of mass.

The eigenmodes for the spinor in the supersymmetric Hilbert space~\eqref{shiftedeigenfunc} have the degree of growth $0$ as they have the following asymptotic behavior, 
\be\ba{l} \label{Fermionasymptshifted}
\psi^+_{\l+ s\frac{\i}{2},k}\sim  {\rm e}^{ \i(k+\frac{1}{2})\theta}  {\rm e}^{-s\i \lambda \eta} \b_{-s\lambda -\frac{\i}{2}} \upsilon_{(s)},
\\
\psi^-_{\l+ s\frac{\i}{2},k} \sim - s\, {\rm e}^{-\i(k+\frac{1}{2})\theta}     {\rm e}^{-s\i \lambda \eta}   \beta_{-s \lambda-\frac{\i}{2}} \upsilon_{(s)}   ,
\ea\ee
where the $\beta_{\l}$ and $\upsilon_{(s)}$ are given in \eqref{alphaupsilon}.

Note that although these eigenmodes survive at the asymptotic boundary of \ads2, being degree of growth~$0$, the inner product among them, defined in~\eqref{innerprodfermion},  is still well defined because the spinorial multiplication of two leading terms in~\eqref{Fermionasymptshifted} vanishes identically due to the projection property of $\upsilon_{(\pm)}$ and only the product of the leading and subleading terms survives, having degree of growth~$-1$.  Furthermore, the inner product~\eqref{innerprodfermion} is  defined in a manner that the parameter $\lambda$ can be complexified. Within this definition 
we find that the supersymmetric basis functions for the spinor field in \eqref{shiftedeigenfunc} form a delta-function orthonormal basis satisfying
\be\ba{l}\label{orthonormalfermionNew}
\langle \psi^\pm_{\lambda + s\frac{\i}{2}, k} | \psi^\pm_{\lambda' + s\frac{\i}{2}, k'} \rangle 
= L^2 \delta(\lambda - \lambda')\delta_{k,k'}\,.
\ea\ee
Here,  the resulting Dirac delta function is natural because we have shifted both of $\lambda$ and $\lambda'$ by same imaginary number $\i/2$.  It may seem that shifting of $\lambda$ and $\lambda'$ by any imaginary number, say $\i\, x$, results in the delta-function in~\eqref{orthonormalfermionNew} while preserving well defined integration. However, we note that the integration in~\eqref{orthonormalfermionNew} is normalizable only if $x \leq 1/2$. To explain, suppose $x>1/2$. Then in the asymptotic behavior~\eqref{Fermionasymptshifted}, the leading term and subleading term behave as $\exp((-1/2+x)\eta)$ and $\exp((-3/2+x)\eta)$ respectively, and thus the integrand in~\eqref{orthonormalfermionNew} has degree of growth $-1 +2x$ which is greater than zero resulting in the integration being divergent.

In contrast to the standard basis given in~\eqref{AdS2eigenfunction} and~\eqref{EigenAdS2a}, the basis functions constructed in~\eqref{shiftedeigenfunc} indeed form a suitable basis to span all the fluctuations that satisfy supersymmetric asymptotic boundary conditions dictated by the variational principle. To demonstrate this, let us exploit the relevant result from the analysis of the supersymmetric boundary condition  in~\cite{GonzalezLezcano:2023cuh} for Euclidean global \ads2. The analysis is reviewed in Appendix~\ref{asympBC} in terms of the convention used in this paper (see also~\cite{Hollands:2006zu,Amsel:2008iz,Correa:2019rdk} for the Lorentzian or Poincare patch of Euclidean  \ads2). For a supersymmetric action, if we demand that the variation of the action around the on-shell saddle vanishes, then
the asymptotic expansion of the scalar and spinor fluctuations are restricted as
\be\ba{ll}\label{Bdrycondition}
\delta \phi = \delta\phi_{(0)}\re^{-\Delta_\phi \eta}+ \cdots, ~~&\Delta_\phi >\half\,,
\\
\delta \psi =\delta\psi_{(0)}\re^{-\Delta_\psi \eta}+ \cdots,~~&\Delta_\psi > 0.
\ea\ee
From this, we note that the lowest bound of the scalings for fluctuation of scalar and spinor fields are  given by $\Delta_\phi =1/2$ and  $\Delta_\psi =0$, and they are actually the degree of  growth of our supersymmetric basis functions for scalar and spinor fields in~\eqref{shiftedeigenfunc} respectively.  This implies that all the possible fluctuations that have the asymptotic scaling above their bound  given in~\eqref{Bdrycondition} can be spanned by the supersymmetric basis. Special attention is required  for the case where $1/2 \geq \Delta_\psi>0$. In this range, the spinor fields are seemingly non-normalizable as they cannot be expanded in terms of the standard basis. However, in terms of our supersymmetric basis with the Euclidean inner product, it is still regarded as normalizable boundary condition and corresponding fluctuations can be expanded using the delta-function normalizable basis. 


\section{One-loop in supersymmetric Hilbert space  }

Now, we want to explore the one-loop partition function of a theory consisting of the scalar and spinor fields  
evaluated in the supersymmetric Hilbert space and compare it with the one that would be evaluated in nonsupersymmetric standard Hilbert space that would satisfy asymptotic boundary condition  different from~\eqref{Bdrycondition}. For this purpose, we shall use the heat kernel method. Since the scalar basis is the same for both of supersymmetric and nonsupersymmetric Hilbert spaces, we can focus only on the contribution of the spinors having the kinetic term $-\i\bar{\psi} (\slashed{D} +M_\psi)\psi$, where $\bar{\psi}$ and $\psi$ are independent spinors. 

A crucial difference between the two cases  is that, while the nonsupersymmetric Hilbert space can suffer from fermionic zero mode, our supersymmetric Hilbert space does not: for the former case, the kinetic operator of the spinor vanishes at $\lambda=0$ when $M_\psi =0$ and the spectrum exists at this point as noticed in~\eqref{spectralF}, whereas for the later case even though the kinetic operator vanishes at $\lambda= 0$ when $M_\psi = -1/2$, such point in the spectrum does not exist as was pointed out in~\eqref{shifteddensity}.  In general, if there are zero modes, we separate out their regularized contribution as $Z_{\text{1-loop}}= Z_{\textrm{zm}}Z'_{\text{1-loop}}$. However, as we do not have zero modes, we can directly compute the one-loop partition function via the standard procedure of heat kernel method \cite{Vassilevich:2003xt} as,
\be\ba{l}\label{HKcalculation}
\log Z^{\psi}_{\text{1-loop}}= \half {\int}_{\epsilon/ L^2 }^\infty \frac{{\rm d }\bar{s}}{\bar{s}} \, K_{\psi}(\bar{s})\,,
\\
K_{\psi}(\bar{s}) \equiv  - \Tr \exp \Bigl[ -\bar{s} \bigl(\i L (\slashed{D} +M_\psi) \bigr)^2\Bigr]
\\
= - 2\int \textrm{d}^2 x \sqrt{g} \int_{\cC} {\rm d}z\, \mu_{\psi}\! (z)   \exp\Bigl[- \bar{s} (z+\i L M_\psi)^2  \Bigr] \,.
\ea\ee
Here we have defined $K_\psi(\bar{s})$  as minus of trace over heat kernel for the Dirac spinors, which is to be integrated over its dimensionless argument $\bar{s}$ from the UV cutoff $\epsilon/L^2$ to infinity. As the trace is over the complete Hilbert space and the quantity $K_\psi(\bar{s})$ is a local quantity.

In the last line of \eqref{HKcalculation}, the trace now involves integration over the complex spectrum parametrized by $z = \lambda +\i /2$.   We note that this is nothing but the heat kernel in standard delta-function normalizable basis with the  contour in $\lambda$ shifted by $\i/2$.
Let us recall that the measure $\mu_\psi(z)$, which is the spectral function, was defined such that it is meromorphic function of $z$ as in~\eqref{spectralF}, and it has poles at $z= \pm \i , \pm 2 \i , ...$. Therefore, the shift of the contour does not cross any poles and thus deos not change the result of the heat kernel. 
This concludes  that the local contribution of the heat kernel method is the same for the supersymmetric and the nonsupersymmetric result. However, the nonlocal contribution  can be different as the standard nonsupersymmetric basis can have fermionic zero mode.

This result explains how  the one-loop study of supersymmetric black holes  via the standard  heat kernel  agrees with the supersymmetric result. For the near horizon geometry of  supersymmetric black holes of the form \ads2$\times$ S$^2$, there are no zero modes of spinor fields since the Dirac operator along S$^2$ does not give zero mass in the Kaluza-Klein tower. Therefore, there is no difference even in the global contribution of heat kernel computation between nonsupersymmetric and supersymmetric Hilbert spaces.

\section{Discussion}  
\indent  
It is worth emphasizing that the quantum fluctuations of fields in supersymmetric theories should reside in a supersymmetric Hilbert space. Therefore, our construction serves as a basic foundation for quantum studies of supersymmetric theories on \ads2, including supersymmetric black hole entropy. We note that in the exact results using supersymmetric localization \cite{Jeon:2018kec,Iliesiu:2022kny,GonzalezLezcano:2023cuh}, the boundary conditions of fields  indeed meet the conditions of the basis functions in our supersymmetric Hilbert space. 

\indent  It is also worth emphasizing that, unlike the standard basis, the spinor basis in~\eqref{shiftedeigenfunc} does not admit fermionic zero modes. Therefore, it ensures  that the path integral is nonvanishing.

\indent Although this paper analyses chiral multiplet fields on \ads2, we expect that the complexified spectrum is pervasively necessary for supersymmetry in generic situations. Specifically, on higher dimensional AdS$_d$, the spectrum of the chiral multiplet will also have  $\i/2$ shift independently of $d$. Other multiplets, such as vector and gravity multiplets should also have a complex spectrum. The explicit construction of the basis for them is an interesting subject for future research.  
 
\indent Even though the new basis functions are delta-function normalizable, the asymptotic behavior shown in~\eqref{Fermionasymptshifted} reveals that the spinor modes  can reach the boundary of \ads2. It would be interesting to investigate the implications of this property of the modes for black hole physics.

\vspace{-2mm}
\section*{Acknowledgements} 
\indent  This work is supported by an appointment to the JRG Program at the APCTP through the Science and Technology Promotion Fund and Lottery Fund of the Korean Government and  by the National Research Foundation of Korea (NRF) grant funded by the Korea government (MSIT) (No. 2021R1F1A1048531). We wish to thank  Chen-Te Ma, Sameer Murthy, Leopoldo A. Pando Zayas, Jeong-Hyuck Park,  Matthew Roberts and Ashoke Sen for useful discussions.



%

\begin{widetext}
\appendix
\section{Special functions}
In this Appendix, we present some useful properties concerning the hypergeometric function  \cite{Gradshteyn:1702455} and gamma function.
\paragraph{Definition. $-$}~The hypergeometric function is defined by the power series
\be
F(\alpha\,,\beta\,; \gamma\,; z)= 1 + \frac{\alpha \cdot \beta}{\gamma \cdot 1}z + \frac{\alpha(\alpha +1)\beta(\beta+1)}{\gamma(\gamma+1)\cdot 1\cdot 2}z^2+ \frac{\alpha(\alpha +1)(\alpha +2)\beta(\beta+1)(\beta+2)}{\gamma(\gamma+1)(\gamma+2)\cdot 1\cdot 2 \cdot 3}z^3 +\cdots\,.
\ee
Integral representation
\be\label{integralrepF}
F(\alpha\,,\beta\,; \gamma\,; z)=\frac{1}{B(\beta\,, \gamma-\beta)}\int_0^1{\rm d}t~ t^{\beta-1 }(1-t)^{\gamma-\beta-1}(1-tz)^{-\alpha}\,,~~~~[{\rm Re}\, \gamma >{\rm Re} \,\beta>0]\,,
\ee
where $B(\alpha\,,\beta)$ is Euler's beta function
\be
B(\alpha\,,\beta)= \frac{\Gamma(\alpha)\Gamma(\beta)}{\Gamma(\alpha+\beta)}\,.\nn
\ee
\paragraph{Useful relations, $-$}~
Inversion formula: 

\begin{eqnarray}\label{Inverse2F1}
&&F(\alpha\,,\beta\,; \gamma\,; z)=\frac{\Gamma(\gamma)\Gamma(\beta-\alpha)}{\Gamma(\beta)\Gamma(\gamma-\alpha)}(-z)^{-\alpha}F\left(\alpha\,,\alpha+1-\gamma\,; \alpha +1 -\beta\,; \frac{1}{z}\right)\nn\\
&&~~~~~~~~~~~~~~~~~~~~~~+\frac{\Gamma(\gamma)\Gamma(\alpha-\beta)}{\Gamma(\alpha)\Gamma(\gamma-\beta)}(-z)^{-\beta}F\left(\beta\,,\beta+1-\gamma\,; \beta +1 -\alpha\,; \frac{1}{z}\right)\,,\\
&&~~~~~~~~~~~~~~~~~~~~~~~~~~~~~~~~~~~~~~~~~[ |\arg z|< \pi\,,~~\alpha-\beta \neq \pm m\,, ~~m= 0,1,2, \cdots ]\,.\nn
\end{eqnarray}
Euler's transformation: 
\begin{eqnarray}\label{Transformular2}
&&F(\alpha\,,\beta\,; \gamma\,; z)=(1-z)^{\gamma- \alpha-\beta}F(\gamma-\alpha\,, \gamma-\beta\,; \gamma\,; z)\,.
\end{eqnarray}
Gauss' recursion relations: 
\be\ba{lll} \label{eq:gaussrecursion}
0&=&(2\beta -\gamma -\beta z+\alpha z)F(\alpha, \beta;\gamma; z)+ (\gamma-\beta)F(\alpha, \beta-1;\gamma;z)
+ \beta(z-1)F(\alpha,\beta+1;\gamma; z)\,,
\\
0&=&\gamma F(\alpha, \beta;\gamma; z)- (\gamma-\beta)F(\alpha, \beta; \gamma+1; z) - \beta F(\alpha, \beta+1; \gamma+1; z) \,,
\\
0&=&\gamma F(\alpha, \beta;\gamma; z)- (\gamma-\alpha)F(\alpha, \beta; \gamma+1; z) - \alpha F(\alpha+1, \beta; \gamma+1; z) \,,
\\
0&=&\gamma F(\alpha,\beta;\gamma;z)-\gamma F(\alpha,\beta+1;\gamma;z)+\alpha z  F(\alpha+1,\beta+1;\gamma+1;z)\, , \\
0&=& \g \left(1-z\right) F(\alpha,\beta;\gamma;z) + \left(\g-\b\right)z F(\alpha,\beta;\gamma+1;z)-\g F(\alpha-1,\beta;\gamma;z) \, ,
\\
0&=& \g \left(1-z\right) F(\alpha,\beta;\gamma;z) + \left(\g-\a\right)z F(\alpha,\beta;\gamma+1;z)-\g F(\alpha,\beta-1;\gamma;z)\,.
\ea\ee
Integration:
\beqa\label{integrationF}
&&\int_0^\infty {\rm d}x\, x^{\gamma -1}(x+z)^{-\sigma} F(\alpha\,, \beta\,; \gamma\,; -x) \\
&&= \frac{\Gamma(\gamma)\Gamma(\alpha - \gamma +\sigma)\Gamma(\beta- \gamma +\sigma)}{\Gamma(\sigma)\Gamma(\alpha +\beta -\gamma +\sigma)} F(\alpha - \gamma +\sigma\,, \beta -\gamma +\sigma\,; \alpha +\beta -\gamma +\sigma \,; 1-z)\,,\nn\\
&&~~~~~~~~~~~~~~~~~~~[ {\rm Re} \,\gamma > 0\,,~~{\rm Re}(\alpha-\gamma +\sigma) > 0\,,~~ {\rm Re}(\beta -\gamma +\sigma)>0\,, ~~|\arg z|< \pi\  ]\,.\nn
\eeqa
\paragraph{Some properties of gamma function.$-$}
Euler's reflection formula:
\be
\Gamma(1-z)\Gamma(z) = \frac{\pi}{\sin\pi z}\,,\qquad z\notin \mathbb{Z}\,.
\ee
Legendre duplication formula:
\be
\Gamma(z)\Gamma\Bigl(z+\frac{1}{2}\Bigr)\= 2^{1-2z} \sqrt{\pi}\,\Gamma(2z)\,.
\ee
Formula for absolute value: 
\be
\left| \Gamma(a+\i b)\right|^2 = \left| \Gamma(a)\right|^2 \prod_{k=0}^{\infty}\frac{1}{1+ \frac{b^2}{(a+k)^2}}\,.\label{Gammauseful1}
\ee
In particular,
\beqa
	&&\left| \Gamma(\frac{1}{2}\pm n  +\i b)\right|^2 = \frac{\pi}{\cosh(\pi b)} \prod_{k=1}^{n}\left( (k-\frac{1}{2})^2 +b^2\right)^{\pm 1}\,,~~n \in \mathbb{N} \,,
	\\
	&&\left|\Gamma(\i\lambda)\right|^2 = \frac{\pi}{\lambda \sinh(\pi \lambda)}\,, ~~~~~|\Gamma(\half +\i\lambda)|^2 = \frac{\pi}{\cosh(\pi \lambda)}\,, \label{Gammauseful3}~~~
|\Gamma(2\i\lambda)|^2
 = \frac{1}{4\pi}\left|\Gamma(\i\lambda)\right|^2|\Gamma(\half +\i\lambda)|^2  \, . \label{Gammauseful4} 
	\eeqa 
	\paragraph{Dirac delta function.$-$} The representation in terms of gamma matrix is given as  
\beqa
 \lim_{\epsilon\rightarrow 0} \frac{\Gamma(-\i \lambda+\epsilon)\Gamma(\i\lambda+\epsilon)}{\Gamma(\epsilon)} = \lim_{\epsilon \rightarrow 0}  \Gamma(\epsilon) \prod_{k=0}^{\infty}\frac{1}{1+\frac{ \lambda^2}{(\epsilon + k)^2}}=  \lim_{\epsilon \rightarrow 0}  \frac{\epsilon}{\lambda^2 +\epsilon^2} \prod_{k=1}^{\infty}\frac{1}{1+\frac{ \lambda^2}{k^2}} 
 =  \label{GammaDeltafn} \lim_{\epsilon \rightarrow 0}  \frac{\epsilon}{\lambda^2 +\epsilon^2}  
   = \pi \delta(\lambda)\,.
\eeqa
Here,  for the first equality we have used the formula \eqref{Gammauseful1}, and for the third equality we have used the fact that  the products  $\prod_{k=1}^{\infty}(\cdots) $ give 1 for $\lambda =0$.

\section{Orthonormality of eigenbasis}\label{Appendix:Innerproduct}
In this Appendix, we  prove  the orthonormality for scalar basis \eqref{AdS2eigenfunction}. This can serve as the building block to show the orthonormality of spinor basis.

The inner product  is given by
\beqa\label{IntegrationI1}
\langle \sc_{\lambda, k} |\sc_{\lambda', k'}\rangle
&=&L^2 \delta_{ k  k'} c_{\lambda, | k|} c_{\lambda' ,| k |} 2\cdot 4^{| k|} I_1\,, \quad 
    I_1  \equiv  \int_0^\infty  {\rm d} x \,x^{| k|}(x+1)^{| k|} 
F(\a\,, \b\,,\gamma\,; -x) F(\alpha'\,, \beta'\,, \gamma'\,; -x)\,,
\eeqa
where we have redefined the variable as $x= \sinh^2 \frac{\eta}{2}$, and have denoted
\be\label{definitions}
c_{\lambda,| k|}\equiv  \frac{1}{2^{| k|}| k|!} \!\left(\! \frac{\Gamma(\frac{1}{2}+| k| +\i\lambda)\Gamma(\frac{1}{2}+| k| -  \i \lambda )}{\Gamma(\i\lambda)\Gamma(-\i\lambda)} \right)^{\!\frac{1}{2}}\,,~~~\alpha \equiv \i\lambda +\half +| k |\,,~~\beta \equiv -\i \lambda +\half +| k | \,,~~~\gamma\equiv | k| +1\,.
\ee
To evaluate the integration $I_1$ in \eqref{IntegrationI1}, we use the transformation formula \eqref{Transformular2} and the integral representation of the hypergeometric function \eqref{integralrepF} to obtain
\begin{align}\label{Integration1}
\begin{split}
I_1 &=  \int_0^\infty  {\rm d} x ~ x^{| k|} 
F(\gamma-\a\,, \gamma- \b\,,\gamma\,; -x) F(\alpha'\,, \beta'\,, \gamma'\,; -x)
= \frac{ 1}{B(\gamma-\b,\b)}\int_0^1 {\rm d}t \,t^{\a-\b-1}(1-t)^{\b-1} I_2(t) \, ,
 \end{split}
\end{align}
where  we define
\begin{align} \label{second}
    I_2(t) & \equiv \int_0^\infty  {\rm d} x \left(\frac{1}{t}+ x\right)^{\a-\gamma} x^{| k|}
 F(\alpha'\,, \beta'\,, \gamma'\,; -x)\,.
\end{align}
Here, we note that the relation \eqref{integralrepF} can be used provided the following constraints are met: 
\begin{align} \label{eq:constraint1}
    \text{Re}(\g) > \text{Re}(\g -\b) >0\, ,
\end{align}
which is always satisfied according to \eqref{definitions}.

Now, to calculate \eqref{second},  we want to apply the integration formula \eqref{integrationF} which  requires the following conditions:
\begin{align} \label{eq:constraints2}
\text{Re}(\a^\prime- \g^\prime +\g - \a) >0\, , \quad \text{Re}(\b^\prime- \g^\prime +\g - \a)>0\,, \quad \text{Re}(\g^\prime) >0\,, \quad \text{arg}\left(\frac{1}{t}\right) < \pi \, ,
\end{align}
where the last two relations in \eqref{eq:constraints2} are always satisfied while the first two, using \eqref{definitions}, reduce to:
\begin{align}
 \text{Re}(\a^\prime - \a) = 0\, , \quad \text{Re}(\b^\prime - \a) = 0\,.
\end{align}
We can avoid violating the bounds by introducing a regulator in \eqref{eq:constraints2} as follows:
\be \label{eq:epsilonReg}
\alpha \rightarrow \alpha +\epsilon \,,~~\beta\rightarrow \beta +\epsilon\,,~~\gamma \rightarrow \gamma +2\epsilon\,,~~~~\epsilon>0\,,
\ee which will guarantee the applicability of \eqref{integrationF} and we take the limit $\epsilon \rightarrow 0$ at the end of the calculation. Hence
\begin{align} \label{I2t}
\begin{split}
    I_2(t) & = \frac{\Gamma(\gamma' ) \Gamma(\alpha'-\gamma' +\gamma -\a +\epsilon )\Gamma(\beta' -\gamma' +\gamma  -\a + \epsilon ) }{\Gamma(\gamma -\beta +\epsilon )\Gamma(\alpha' +\beta' -\gamma' +\gamma -\a  +\epsilon )} \\
    & \times  F(\alpha' -\gamma' +\gamma -\a +\epsilon 
\,, \beta' -\gamma'+\gamma -\a +\epsilon  \,,  \alpha' +\beta' -\gamma' +\gamma-\a +\epsilon \,; 1-\frac{1}{t})\, . 
\end{split}
\end{align}

We are now left with  the integration given in \eqref{Integration1}, 
\beqa\label{integ1}
I_1
&=& \frac{ 1}{B(\gamma-\b,\b)} \int_0^1 {\rm d}t \,t^{\a-\b-1}(1-t)^{\b-1 +\epsilon } 
 I_2(t)\,.
\eeqa
We redefine the variable 
$
1-\frac{1}{t}=- x\,,
$
and use the integration formula \eqref{integrationF} again. Then
\begin{align}
\begin{split}
I_1
&= \frac{ 1}{B(\gamma-\b,\b)}
\int_0^\infty {\rm d}x ~  (1+x)^{-(\a + \epsilon )} \,x^{\b-1 + \epsilon  } I_2(x) \\
&= \frac{ 1}{B(\gamma-\b,\b)} \frac{\Gamma(\gamma' ) \Gamma(\alpha'-\gamma' +\gamma -\a +\epsilon )\Gamma(\beta' -\gamma' +\gamma  -\a + \epsilon )\Gamma(\beta-\beta' +\epsilon )\Gamma(\beta-\alpha' +\epsilon ) }{\Gamma(\gamma -\beta +\epsilon )\Gamma(\alpha' +\beta' -\gamma' +\gamma -\a  +\epsilon )\Gamma(\beta +\epsilon )\Gamma(\gamma-\gamma' +2 \epsilon )}\,,\label{integ2}
\end{split}
\end{align}
where we can check that the conditions in \eqref{integrationF} are satisfied due to the regularization parameter $\epsilon$ as
\beqa 
{\rm Re} \left( \beta -\beta' + \epsilon \right)>0\,, \quad {\rm Re} \left( \beta -\alpha' + \epsilon \right)>0\,.  
\eeqa
By taking the $\epsilon \rightarrow 0$ and recovering all the factors defined in \eqref{definitions}, we obtain
\beqa
\langle \sc_{\lambda, k} |\sc_{\lambda', k'}\rangle
&=&  L^2 2\delta_{ k  k'} \frac{1}{|\Gamma(\i\lambda)|^2|\Gamma(\half +\i \lambda)|^2} \frac{|\Gamma(\i(\lambda +\lambda')+\epsilon)|^2 |\Gamma(\i (\lambda -\lambda')+\epsilon)|^2}{\Gamma(2\epsilon)}\nn
\\ \label{eq:twodeltas}
&=& L^2 4 \pi \delta_{ k  k'} \frac{|\Gamma(\i 2\lambda)|^2}{|\Gamma(\i\lambda)|^2|\Gamma(\half +\i \lambda)|^2}\left(\delta(\lambda-\lambda' )+ \delta(\lambda+\lambda')\right)
\\
&=&  L^2\delta_{ k  k'} \left(\delta(\lambda-\lambda' )+ \delta(\lambda+\lambda')\right)\,,\nn
\eeqa
where  we use the identity \eqref{GammaDeltafn} for the second equality  and \eqref{Gammauseful4} for the third equality. The fact that $\lambda > 0$ and $\lambda' > 0$ for scalar basis yields the orthonormal condition \eqref{scalarinnerprod}.

\section{Asymptotic boundary condition}\label{asympBC}


For the chiral multiplet 
on \ads2 and $s=+1$ in~\eqref{KSEAdS2}, the supersymmetric Lagrangian is given by 
\be\ba{l}
\label{Actionchiral}
\cL  = \partial_{\mu}\overline{\phi}\,\partial^{\mu}\phi  
+ M_\phi^2   \overline{\phi}\phi
 \!- \i \overline{\psi}( \slashed{D} +\! M_\psi) \psi   + \nabla_\mu V^\mu   ,
\\
 V^{\mu}\equiv \frac{1}{2\i \Bar{\epsilon}\epsilon} \bigl[\i \epsilon^{\mu \nu}   (\i \bar \epsilon \gamma_3 \epsilon)    \left(\bar \phi \,\partial_\nu \phi -  \phi \,\partial_\nu \bar\phi \right)  
- (\bar\epsilon \gamma^\mu \bar \psi ) ~ \epsilon  \psi + (\epsilon \bar \psi ) \bar \epsilon  \gamma^\mu  \psi \bigr]\,,
\ea\ee
with mass relation
$(L M_\phi)^2 = (L M_\psi)^2 + L {M_\psi},$~thereby satisfying the Breitenlohner-Freedman bound~\cite{Breitenlohner:1982jf,Breitenlohner:1982bm}, $(LM_\phi)^2 \geq -1/4 $. 
Here, the total derivative term is chosen \cite{GonzalezLezcano:2023cuh}
such that it  ensures the total Lagrangian is invariant under the following transformations:
\be
\ba{l}\label{deltachiral}
 Q \phi   = \bar\epsilon \psi \, , \quad \quad Q \psi  = \i \gamma^\mu \epsilon\, \partial_\mu \phi  
-\i \epsilon M_\psi \phi \,,
\\
Q \bar\phi  = \epsilon \bar\psi \, ,   \quad \quad  Q \bar\psi  = \i \gamma^\mu \bar\epsilon \,\partial_\mu \bar\phi  
-\i \bar \epsilon M_\psi \bar\phi\,. 
\ea
\ee 
We note that two spinor fields $\psi$ and $\overline{\psi}$ are independent spinors in Euclidean theory.
Likewise, $\epsilon$ and $\overline{\epsilon}$ are also independent Killing spinors each of which can \textit{a priori} be spanned by two solutions~\eqref{eq:KillingSpinors}.
However, since the presence of boundary preserves only half of the supersymmetries, we set $\epsilon = \varepsilon^+$ and $\overline{\epsilon}= \varepsilon^-$ as the preserved supersymmetry such that  the Killing vector $\bar{\epsilon}\gamma^\mu \epsilon \partial_\mu = \partial_\theta$ preserves the boundary of \ads2.

The  asymptotic  
expansion 
of the fields is determined by asymptotic equations of motion with two expansion coefficients. For the scalar and the Dirac spinor, we have respectively 
\be\ba{ll}\label{FGexpansion}
\phi\! = \phi^{+}_{(0)} {\rm e}^{- \Delta^{+}_\phi \eta} 
+\! \phi^{-}_{(0)} {\rm e}^{- \Delta^{-}_\phi \eta} +\cdots ,  & \Delta^{\pm}_\phi  = \frac{1}{2} \pm
 \sqrt{\frac{1}{4} +(L M_{\phi})^2}= \half \pm \bigl|LM_\psi +\half\bigr|\,,
\\
\psi  = \psi_{(0)}^{+} \upsilon_{(+)}{ \rm e}^{- \Delta^{+}_\psi \eta} 
+ \psi_{(0)}^{-} \upsilon_{(-)}{ \rm e}^{- \Delta^{-}_\psi \eta} +\cdots\,,~~&\Delta_{\psi}^{\pm}      = \frac{1}{2} \pm L M_\psi\,.
\ea
\ee
The same expansion applies to $\overline{\phi}$ and $\overline{\psi}$, so we omit their analysis.

Since supersymmetry relates the masses of scalar and spinor fields, it also relates the expansion coefficients and the corresponding scaling behavior of scalar and spinor in \eqref{FGexpansion}, depending on the range of fermionic mass as
\be\ba{lll}
Q \phi_{(0)}^{\pm} \!= \psi_{(0)}^{\pm}, &\Delta^{\pm}_\phi \!= \Delta^{\pm }_\psi \pm\frac{1}{2}, &\mbox{for } M_\psi \geq -\half,
\\
Q \phi_{(0)}^{\pm} \!= \psi_{(0)}^{\mp}, &\Delta^{\pm}_\phi \!= \Delta^{\mp}_\psi \mp\frac{1}{2}, &\mbox{for }M_\psi <-\half.
\ea\ee

The variational principle determines the boundary condition for the expansion coefficients as well as the allowed range of the  fall-off  of quantum fluctuations. If we insert the  asymptotic  expansion~\eqref{FGexpansion} into the variation of action~\eqref{Actionchiral} and demand that it vanishes on-shell, then we can obtain: 
\be\ba{lll}\label{Bdrycondition}
0= \phi_{(0)}^{-}=\delta \phi_{(0)}^{-} = \psi^{-}_{(0)} = \delta \psi^{-}_{(0)}\,, \;\;&
\Delta_{\delta\phi}>\half\,,\;\; \Delta_{\delta\psi} >0\,,~~~~~&\mbox{for  }M_\psi >-\half,
\\
0= \phi_{(0)}^{-}=\delta \phi_{(0)}^{-} = \psi^{+}_{(0)} = \delta \psi^{+}_{(0)}\,, &
\Delta_{\delta\phi}>\half\,, \;\;\Delta_{\delta\psi} >1\,, ~&\mbox{for }M_\psi\! <\!-\half \,\,\text{and }\, M_\psi \!\neq\! -1\,,
\ea\ee
where $\phi^{-}_{(0)}\neq 0$ is allowed for the case where $M_\psi =-1$.

	
\end{widetext}

\end{document}